\documentclass[showpacs,eqsecnum,aps]{revtex4}
\usepackage{amsmath}
\usepackage{amssymb}
\usepackage{amsfonts}
\usepackage{graphicx,color}
\def\beq{\begin{equation}}
\def\eeq{\end{equation}}
\def\beqarray{\begin{eqnarray}}
\def\eeqarray{\end{eqnarray}}

\def\<{{\langle}}
\def\>{{\rangle}}
\def\q{{\mathbf q}}

\def\B{{\mathbf B}}
\def\p{{\mathbf p}}
\def\P{{\mathbf P}}
\def\v{{\mathbf v}}
\def\V{{\mathbf V}}
\def\B{{\mathbf B}}
\def\b{{\mathbf b}}

\def\k{{\mathbf k}}

\def\fq{{\tilde{\mathbf q}}}

\def\fp{{\tilde{\mathbf p}}}
\def\fP{{\tilde{\mathbf P}}}
\def\fB{{\tilde{\mathbf B}}}
\def\fb{{\tilde{\mathbf b}}}
\def\F{{\cal F}}
\def\half{{\frac{1}{2}}}
\usepackage{graphicx,color}

\begin{document}

\title{Model Tests of Cluster Separability In Relativistic Quantum Mechanics}

\author{B. D. Keister}
\affiliation{
Physics Division, National Science Foundation, Arlington, VA 22230
}

\author{W. N. Polyzou}
\affiliation{
Department of Physics and Astronomy, The University of Iowa, Iowa City, IA
52242}

\vspace{10mm}

\date{\today}

\begin{abstract}

  A relativistically invariant quantum theory first advanced by
  Bakamjian and Thomas has proven very useful in modeling few-body
  systems.  For three particles or more, this approach is known
  formally to fail the constraint of cluster separability, whereby
  symmetries and conservation laws that hold for a system of particles
  also hold for isolated subsystems.  Cluster separability can be
  restored by means of a recursive construction using unitary
  transformations, but implementation is difficult in practice, and
  the quantitative extent to which the Bakamjian-Thomas approach
  violates cluster separability has never been tested.  This paper
  provides such a test by means of a model of a scalar probe in a
  three-particle system for which (1) it is simple enough that there
  is a straightforward solution that satisfies Poincar\'e invariance
  and cluster separability, and (2) one can also apply the
  Bakamjian-Thomas approach.  The difference between these
  calculations provides a measure of the size of the corrections from
  the Sokolov construction that are needed to restore cluster
  properties.  Our estimates suggest that, in models based on nucleon
  degrees of freedom, the corrections that restore cluster properties
  are too small to effect calculations of observables.

\end{abstract}

\vspace{10mm}


\maketitle

\section{Introduction}

There are two distinct requirements for describing quantum mechanical
systems of particles under the requirements of special relativity.
The first requirement is Poincar\'e invariance: probabilities,
expectation values and ensemble averages for equivalent experiments
performed in different inertial frames are identical. A necessary and
sufficient condition for a quantum theory to be Poincar\'e invariant
is that the dynamics is described by a unitary representation of the
Poincar\'e group \cite{Wigner:1939cj}.  The second requirement is
cluster separability: isolated subsystems must have the same
observable properties as they would in a framework in which the other
``spectator'' particles are absent entirely.  This requirement
justifies tests of special relativity on isolated subsystems.  It
applies both to systems of particles interacting among themselves
(e.g.~via the strong interaction) and to the current operators that
allow them to interact with external fields.

There is more than one way to implement the requirements of Poincar\'e
invariance in a quantum system, and cluster separability is not
necessarily an automatic consequence of the implementation.  In local
quantum field theory, Poincar\'e invariance and cluster separability
are satisfied formally as a consequence of the covariance, spectral
properties and the local commutation relations of the field operators.
For systems amenable to perturbation methods, these properties can
still hold.  For non-perturbative systems (such as those based on the
strong interaction) controlled approximations are replaced by a
truncation or an expansion scheme, which replaces an infinite
hierarchy of coupled non-linear field equations with a more tractable
finite subset, and which may be valid only in restricted parameter
domains.  These schemes may still exhibit Lorentz covariance, but
cluster separability does not automatically follow, and must be
demonstrated on a case-by-case basis that depends on the scheme.

An alternative implementation of Poincar\'e invariance in a few-body
system involves direct construction of the dynamical representations
of the Poincar\'e group in the presence interactions with a finite
number of particle (rather than field) degrees of freedom.  Such a
construction was originally studied by Dirac~\cite{Dirac:1949cp}, and
Bakamjian and Thomas~\cite{Bakamjian} provided an explicit
construction of the generators for a system of two interacting
particles.

Many realistic, Poincar\'e invariant, quantum mechanical models of
strongly interacting few-body systems are based on a generalization of
the Bakamjian-Thomas construction \cite{Keister:1991sb}.  This
construction can be used to construct models of systems of arbitrary
numbers of particles and systems that do not conserve particle number.
A representative sample of few-body applications of Poincar\'e
invariant quantum mechanics based on the Bakamjian-Thomas construction
includes relativistic constituent quark models
\cite{Melde:2008yr,Krutov:1999zz,Coester:2005cv,Sengbusch:2004sf,Boffi:2001zb},
relativistic few-nucleon models
\cite{Lin:2007kg,Witala:2004pv,Witala:2005nw,Witala:2008va,Witala:2011yq},
relativistic models involving electromagnetic probes
\cite{Huang:2008jd,Keister:1996ny,Keister:1988zz}, and relativistic
models with particle production \cite{Fuda:2009zz,Kamano:2008gr}.

The virtue of the Bakamjian-Thomas construction is that it provides a
means for constructing Poincar\'e generators for systems of
interacting particles, and where the problems have a finite number of
degrees of freedom and are in principle solvable.  The limitation of
this construction is that for systems of more than two particles, the
dynamical representation of the Poincar\'e group does not become a
tensor product on states representing asymptotically separated
subsystems.  While this limitation does not lead to observable
consequences in the three-body $S$-matrix or bound-state observables,
there are observable consequences when the three-body system is
embedded in the four-particle Hilbert space, as it is in four-body
problems.

The most ambitious applications of the Bakamjian-Thomas construction
are relativistic Faddeev
calculations~\cite{Lin:2007kg,Witala:2004pv,Witala:2005nw,Witala:2008va,Witala:2011yq}
of three-nucleon scattering, which have been performed using realistic
nucleon-nucleon forces and three-nucleon forces.  The next step in
developing a relativistic few-body theory would be to model
four-nucleon systems or electron scattering from a three-nucleon
system.  In both of these applications a three-nucleon subsystem is
embedded in a four-particle Hilbert space, and, for the first time,
there is the possibility of observable consequences of violations of
cluster properties.

The restoration of cluster properties can be achieved through a
recursive construction due to Sokolov
\cite{Sokolov:1977,Coester:1982vt}.  The Sokolov construction
generates a dynamical representation of the Poincar\'e group
satisfying cluster properties using of a hierarchy of unitary
transformations, each of which preserves the $S$ matrix while
transforming tensor products of subsystem representations of the
Poincar\'e group to representations where the interactions can be
added in a manner that preserves the underlying Poincar\'e symmetry.
The Bakamjian-Thomas representation is retrieved by setting the
relevant unitary transformations to the identity operator.  The
Sokolov construction is non-trivial and the unitary transformations
depend upon the interactions.  An isolated three-body system is a
special case where the cluster property can be achieved by means of a
single overall unitary transformation that preserves the $S$ matrix.
For systems of more than three particles there are observable
differences between the $S$ matrices in the two representations.  The
interested reader can find a complete discussion in
\cite{Coester:1982vt}.

The construction of the Sokolov hierarchy of unitary transformations
is sufficiently complicated that the technique has never been used in
realistic calculations.  Before undertaking computationally intensive
four-body calculations, one would like to know the importance of the
corrections required by cluster properties.  While
Ref.~\cite{Coester:1982vt} argued that these corrections should be
small in nuclear physics applications, this was never quantified in
any model calculations.  An additional investigation is needed to
determine if these corrections can be ignored, can be treated
perturbatively, or must be treated exactly.  To address this question
we construct a model involving matrix elements of a scalar probe for a
system of three particles.  The model is simple enough that both
Poincar\'e invariance and cluster separability are easily satisfied;
this unusual pair of features is due to the simplicity of the test
model.  At the same time the model is sufficiently rich that it
permits an equivalent Bakamjian-Thomas treatment of the three-body
system that illustrates the quantitative impact of the breakdown of
cluster separability in this four-body problem.  The conclusion of our
preliminary analysis suggests that the corrections required by cluster
properties are too small to be observable in nuclear physics
applications.  The corrections are more important for models based on
sub-nucleon degrees of freedom.

\section{Cluster Properties}

In a few-body quantum mechanical model the requirement of cluster
separability means that the Poincar\'e generators
$\{H,\mathbf{P},\mathbf{J},\mathbf{K} \}$ that generate time
translations, translations, rotations, and rotationless Lorentz
transformations and current operators $I^{\mu} (x)$ have cluster
expansions of the form
\beq 
H= \sum_i H_i + \sum_{ij} H_{ij} + \sum_{jk} H_{ijk} + \cdots
\eeq
\beq
\mathbf{P} = \sum_i \mathbf{P}_i + \sum_{ij} \mathbf{P}_{ij} + \sum_{ijk} 
\mathbf{P}_{ijk} + \cdots
\eeq
\beq
\mathbf{J} = \sum_i \mathbf{J}_i + \sum_{ij} \mathbf{J}_{ij} + \sum_{ijk} 
\mathbf{J}_{ijk} + \cdots
\eeq
\beq
\mathbf{K} = \sum_i \mathbf{K}_i + \sum_{ij} \mathbf{K}_{ij} + \sum_{ijk} 
\mathbf{K}_{ijk} + \cdots
\eeq
\beq 
I^{\mu}(x)= \sum_i I^{\mu}_i(x) + \sum_{ij} I^{\mu}_{ij}(x) + 
\sum_{jk} I^{\mu}_{ijk}(x) + \cdots
\label{eq:b1}
\eeq
where the terms $X_{ij}, X_{ijk},\cdots$ are short-range
$2,3\cdots$-body operators (which only depend on the degrees of
freedom of the particles in a cluster and vanish when one particle in
a cluster is separated from the remaining sub-cluster), and the sums
of the generators and currents over any subset of particles satisfies
the Poincar\'e commutation relations, current covariance, and current
conservation.

These constraints are realized in the Sokolov construction, but
they are not generally satisfied by the Bakamjian-Thomas construction.
The Sokolov construction generates all of the many-body
operators that appear in these cluster expansions as functions of
input two-body interactions.   These interactions are necessary to 
preserve the Poincar\'e commutation relations in the presence of 
interactions.

\section{Test Model}

Poincar\'e invariance and the Dirac forms of dynamics for
Bakamian-Thomas ($BT$) constructions are discussed extensively in
Ref.~\cite{Keister:1991sb}.  The framework provided here makes use of
that discussion.  Conceptually, the Bakamjian-Thomas construction
defines the dynamics in the rest frame of the non-interacting system
and then uses simultaneous eigenstates of the mass and spin (Casimir
operators for the Poincar\'e group) to construct unitary
representations of the Poincar\'e group.  We approach the model in a
heuristic way in order to illustrate the issue of cluster separability
combined with Poincar\'e invariance with a minimum of formal
development.

A simple four-body model where we can observe the breakdown of cluster
separability consists of a system of three spinless particles
interacting with an external probe that represents the presence of a
fourth particle.  The relevant dynamical quantities are matrix
elements of a one-body scalar density between eigenstates of the
three-body four-momentum operator.  We will see that, even for this
simple model, there are differences in matrix elements of the scalar
density calculated using three-body eigenstates of a $BT$
four-momentum operator compared to matrix elements calculated using
three-body eigenstates of a four-momentum operator that clusters.  Our
simplifying assumptions show that the violation of cluster properties
has nothing to do with particle spins, or properties of more realistic
probes like four-current densities.

\begin{figure*}[h]
\includegraphics[height=5cm]{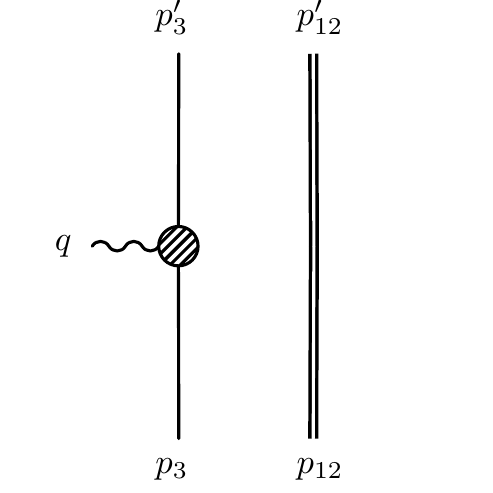}
    \caption{Graph of one-body density interacting with a particle 
plus a bound two-body spectator.}
    \label{fig:graph}
\end{figure*}

The scalar density operator, $j(x)$, is assumed to be a one-body
operator.  The three-particle system consists of a bound pair (which
we label the (12) subsystem) and a third particle (3 which we call the
struck particle) that does not interact with particles 1 or 2.  Only
particle 3 interacts with the scalar density $j(x)$.
Therefore, the (12) subsystem acts as a spectator with respect to the
action of the probe.  This can be illustrated by
disconnected graph shown in Fig.~\ref{fig:graph}.

In the following sections we will calculate matrix elements of the
density operator in this simple system using the Bakamjian-Thomas
unitary representation of the Poincar\'e group, and compare them with
a result that exhibits manifest cluster separability.  The latter
result is possible only because of the simplicity of the model.

\subsection{Tensor-Product ($TP$) Representation}

For the simple system introduced above, 
the unitary representation of the Poincar\'e group is a
tensor product of a two-body representation on a two-particle Hilbert
space and a one-particle representation acting on a one-body Hilbert
space.  In this section we consider a three-body eigenstate that is a
tensor product of bound (12) eigenstate and a one-particle
eigenstate
\begin{equation}
  \label{eq:3a}
  | \lambda, m ; \p_{12}, \p_3 \>_{TP} = | \lambda; \p_{12} \>
\otimes| m,\p_3  \> .
\end{equation}
where $\lambda$ labels the mass eigenvalue $M_\lambda$ of the bound
state of the interacting two-body mass operator, $M_{12}$, $\p_{12}$
is its three-momentum, and $m$ and $\p_3$ are the mass and momentum of
particle 3.  Since the scalar density $j(x)$ acts only in the space of
particle 3, the matrix element has the form
\begin{equation}
  \label{eq:m5}
 _{TP} \< \lambda , m; \p_{12}', \p_3' | j(0) | \lambda , m; \p_{12} ,\p_3 \>_{TP}
  = \delta(\p_{12}' - \p_{12})   \<  m;  \p_3' | j(0) |  m;  \p_3 \>,
\end{equation}
where
\begin{equation}
  \label{eq:m5a}
\sqrt{\omega_m (\mathbf{p}_3')}   \<  m,  \p_3' | j(0) |  m,  \p_3 \>
\sqrt{\omega_m (\mathbf{p}_3)}  = m G(q^2); \quad q = p_3' - p_3,
\end{equation}
and $q$ is the four-momentum transferred to particle 3 and $\omega_m
(\mathbf{p}_3) = \sqrt{m^2 + \mathbf{p}_3^2}$.  The square roots in
Eq.~(\ref{eq:m5a}) ensure the Lorentz invariance of the form factor
$G(q^2)$ when the states of particle 3 have a delta-function normalization:
\begin{equation}
\label{eq:m5cc}
\< m,  \p_3' |  m, \p_3 \> = \delta(\p_{3}' - \p_{3})  .
\end{equation}

If we specify initial momenta, $\p_3$ and $\p_{12}$, and the momentum
transfer, $q$, and integrate the matrix element over the final
spectator momentum, $\p_{12}'$, the integral collapses due to the
spectator momentum delta function in Eq.~(\ref{eq:m5}), and we have
\begin{equation}
  \label{eq:m6}
  \F_{TP} := \int d\p_{12}' 
\sqrt{\omega_m (\mathbf{p}_3')}{}_{TP}\< \lambda , m; \p_{12}', \p_3' | j(0) |
  \lambda , m; \p_{12}, \p_3 \>_{TP} \sqrt{\omega_m (\mathbf{p}_3)} 
 = m G(q^2).
\end{equation}
Equation~(\ref{eq:m6}) is the Tensor-Product ($TP$) result.  It has no
dependence upon the momentum $\p_{12}$ of the bound-state spectator,
as expected from the physical requirement of cluster separability, and
does not depend upon the specific values of $\p_3$ and $\p_3'$, as
long as $p_3' - p_3 = q$.  The result is just the invariant form
factor of particle 3 multiplied by its mass.

The representation for a three-particle system becomes a tensor
product of a one-body and two-body representation of the Poincar\'e
group whenever the interactions involving one of the particles is set to
zero.  This implies that the generators have the form (\ref{eq:b1}).
Ideally one would simply perform all calculations using the $TP$
representation.  However, for examples beyond the simple model
presented here, it is a highly non-trivial task to obtain a
tensor-product representation that preserves Poincar\'e invariance.
The only known method for systems with a finite number of degrees of
freedom is to obtain first the Bakamjian-Thomas representation (which
has interacting generators that satisfy the Poincar\'e commutation
relations), and then to restore cluster separability by a recursive
construction involving a hierarchy Sokolov's unitary transformations.
In the special case of the three-body system the two representations
are related by a single unitary transformation, $A$:
\beq 
\vert \psi \rangle_{TP} = A \vert \psi \rangle_{BT} .  
\label{eq:1a}
\eeq 
The operator $A$ is interaction dependent and contains the corrections
that restore cluster properties to the $BT$ representation.  The
three-body $TP$ and $BT$ representations related by $A$ have the same
3-body $S$ matrix, but off-shell differences lead to different
predictions in the four-body system.  The operator $A$ converts a $BT$
representation to a $TP$ representation satisfying the constraints
imposed by the Poincar\'e commutation relations.  However, for
realistic three-body systems, $A$ is a complicated
interaction-dependent operator.

The size of the difference 
\beq
_{TP}\langle \psi' \vert j(0) \vert \psi \rangle_{TP} -
_{BT}\langle \psi' \vert  j(0)  \vert \psi \rangle_{BT},
\label{eq:2a}
\eeq
where the $BT$ and $TP$ states are related by Eq.~(\ref{eq:1a}), 
provides a measure of the importance of the corrections needed to
restore cluster separability.

While in general the Sokolov operators $A$ need to be computed, the
test model presented here is simple enough that the $TP$ and $BT$
states related by $A$ can be computed directly, and then used in
(\ref{eq:2a}) to estimate the size of the corrections needed to
restore cluster separability.

\subsection{$BT$ Representations}

In the following sections, we evaluate matrix elements of the scalar density
(\ref{eq:m5}) using $BT$ representations of the 2+1-particle system
corresponding to different forms of dynamics~\cite{Dirac:1949cp} and
compare these results to the tensor product result.  These matrix
elements will each depend upon the momentum dependence and mass
eigenvalue of the two-body bound-state wave function, in violation of
cluster separability, and our goal is to examine the magnitude of that
violation.  In all cases, we specify the four momentum $q$ transferred
to the struck particle and bound-state momentum $\p_{12}$ for a given
calculation and then vary these momenta for sensitivity tests.
Dependence of this matrix element on $\p_{12}$ indicates a failure of
cluster properties.

In addition to the external momenta $\p_3$ and $\p_{12}$, the
calculations depend upon the momenta of the constituent particles,
which we label $\b_1$ , $\b_2$ and $\b_3$.  The constituent momenta
are not experimentally accessible, and their relation to the external
momenta depends in turn upon the form of dynamics, as is discussed in
detail below.  For purposes of comparison we vary the external momenta
keeping $\b_3'-\b_3=\q$.  This requires different
kinematic conditions for each form of dynamics.

\subsection{$BT$ Representation: Instant Form}

We now evaluate the scalar density matrix element in Eq.~(\ref{eq:m5}) using 
an instant-form $BT$ model.  Instant-form models have no
interactions in the generators of rotations and space translations. 

We provide first an observation about frames that will be relevant to
each of the $BT$ representations.  Calculations of current-matrix
observables typically use a Breit frame (in which the energy transfer
is zero), a lab frame (initial target three-momentum is zero), or an
anti-lab frame (final target momentum is zero).  There is no frame in
which both initial and final target three-momenta can be zero, so the
calculation will require Lorentz transformations relating frames
with different target momentum.  These are provided explicitly where
needed below.

First, we change variables, replacing the bound state and particle~3
momenta by the total momentum, $\P$, of the system and the momentum of
particle~3 transformed to the frame in which $\P=0$ by a rotationless 
Lorentz transformation $\Lambda^{-1}(\mathbf{P}/M)$:
\begin{equation}
  \label{eq:b1a}   
  (\p_{12}, \p_3) \to (\P, \p),
\end{equation}
where 
\begin{equation}
  \label{eq:b2}
  \P  := 
 \p_{12} + \p_3;\quad 
p :=  \Lambda^{-1} (\mathbf{P}/M)p_3; 
\quad
\p = \p_3 + \Phi_-(\p_3,\P,M)\P,
\end{equation}
and  
\begin{eqnarray}
  \label{eq:b3}
  \Phi_-(\p_3,\P,M) &:=&\frac {1}{ M} 
  \left[ \frac {\P\cdot\p_3}{ E + M} - \omega_m(\p_3)\right];
  \nonumber \\ 
M &=&\sqrt{E^2 - \P^2}; \quad
E = \sqrt{m^2 + \p_3^2} + \sqrt{M_\lambda^2 + \p_{12}^2};
\nonumber \\
\omega_m(\p_3) &=& \sqrt{m^2+\p_3^2}.
\end{eqnarray}
With this variable change the relation between the $(12)$ bound state 
and particle~3 in these bases is:
\begin{equation}
  \label{eq:b4}
  | \lambda , m; \p_{12}, \p_3 \> 
  = \left| \frac{\partial(\P ,\p)}{\partial(\p_{12},\p_3)}\right|^\half_\lambda
  | \lambda ,m; \P, \p \>,
\end{equation}
where the suffix $\lambda$ in the Jacobian indicates that the
interacting two-body mass eigenvalue $M_\lambda$ was used in the transformation.

Equations~(\ref{eq:b1a}-\ref{eq:b4}) connect the momentum variables
via a Lorentz transformation that depends upon the {\it interacting}
mass $M_\lambda$.  The $BT$ construction also uses a transformation
that uses a {\it non-interacting} mass.  To make this connection, we
introduce constituent-particle momenta $\b_1, \b_2, \b_3 $.  We use a
different notation for these constituent momenta because they are not
necessarily the same as $\p_1, \p_2, \p_3$; for example, we will see
that $\b_3 \not= \p_3$.  We define variables analogous to
Eqs.~(\ref{eq:b1a}-\ref{eq:b3}) for the non-interacting system
\begin{equation}
\label{eq:b1b}   
 (\b_{1}, \b_3,\p_3) \to (\B, \b ,\k).
\end{equation}
The variables $\B$ and $\b$ are defined analogously to the definitions
of $\P$ and $\p$:
\beq
\mathbf{B} := \b_{12} +  \b_3 \qquad \mathbf{b}_{12} := \b_1+\b_2 
\eeq
\beq
b := \Lambda^{-1} (\mathbf{B}/M_0)b_3; 
\quad
\b = \b_3
 + \Phi_-(\b_3,\B,M_0)\B,
\label{eq:b7a}
\eeq
\beq
E_0 := \sum_i \sqrt{\b_i^2 + m^2} \qquad M_0 := \sqrt{E_0^2 - \mathbf{B}^2} .
\eeq

The additional variable $\mathbf{k}$ is obtained by (1) first boosting
all three $b_i$ to the rest frame of the non-interacting three-body
system, followed by a boost of particle 1 to the rest frame to the
(12) pair.  The vector $\k$ is the three-momentum of particle 1 after
applying these two transformations.  While the expression for
$\mathbf{k}$ as a function of the momenta $\mathbf{b}_i$ can be
written down explicitly, only the magnitude of $k$ plays a non-trivial
role in our calculations.  It is related to the invariant mass of the
non-interacting two and three-body systems by
\beq
M_k: = 2\sqrt{m^2 + \k^2}
\qquad M_0= \sqrt{M_k^2+\mathbf{b}^2} + \sqrt{m^2+\mathbf{b}^2}.
\label{eq:b4c}
\eeq
%
%

In the instant-form Bakamjian-Thomas construction, we identify
\beq
\B = \P; \quad \b = \p.
\label{eq:b5}
\eeq
The first of these identifications is a general property of an instant-form 
dynamics, while the second equation is a requirement of the Bakamjian-Thomas 
construction.  
 
These identifications and the above definitions imply relations
between the observable momenta $\mathbf{p}_3 , \mathbf{p}_{12}$ and
the constituent-particle momenta, $\b_1,\b_2,\b_3$.  It is because
$\p$ and $\b$ are related to the momentum of particle 3 by Lorentz
boosts involving different masses that $\p_3 \not=\b_3$.

%
The BT four-momentum eigenstate with particles 1 and 2 bound in the
three-particle basis is
\begin{equation}
  \< \B, \b, \k \vert \lambda; \P', \p'  \>_{BT} =
  \delta (\P'-\B) \delta (\p' - \b) 
  \phi_\lambda (\k),
 \label{eq:7c}
\end{equation}
where $\phi_\lambda (\k)$ is the bound-state wave function.  Note that
$\b$ depends upon the free two-particle invariant mass $M_{120}$ via
Eq.~(\ref{eq:b7a}-\ref{eq:b4c}), while $\p'$ depends upon the bound
state mass eigenvalue via $\lambda$ Eq.~(\ref{eq:b2}-\ref{eq:b3}).
The association of $\p$ and $\b$ in the delta function of
Eq.~(\ref{eq:7c}) is only consistent when $\mathbf{P}=0$.  This is
because in the instant form Bakamjian-Thomas construction the
dynamical model is solved in the rest frame and the resulting mass
eigenvalue determines how the system behaves under Lorentz boosts.
The identification of $\p$ with $\b$ becomes inconsistent in frames
where $\P\not= 0$ because these quantities are defined using different
boosts.  Since the initial and final states in a scalar density matrix
element are in different frames, this inconsistency cannot be avoided
and it results in the violation of cluster properties.
 
We now make use of Eq.~(\ref{eq:7c}) to compute the $BT$ counterpart to
$\F_{TP}$ that was defined by Eq.~(\ref{eq:m6}):
\begin{eqnarray}
  \label{eq:b11}
  \F_{BT} &:=& \int d\p_{12}' 
\sqrt{\omega_m (\mathbf{p}_3')}{}_{BT}\< \lambda , m; \p_{12}', \p_3' | j(0) |
  \lambda , m; \p_{12}, \p_3 \>_{BT} \sqrt{\omega_m (\mathbf{p}_3)} 
\\ \nonumber
  &:=& \int   d\p_{12}'
  d\k' 
  d\k 
 {\sqrt{\omega_m (\mathbf{p}_3')}}    
  \\ \nonumber
  &&\quad\times
  \left| \frac{\partial(\P', \p')}{\partial(\p_{12}',\p_3')}\right|^\half_\lambda
  \left| \frac{\partial(\P, \p)}{\partial(\p_{12},\p_3)}\right|^\half_\lambda
  \left| \frac{\partial(\b_{12}',\b_3')}{\partial(\P', \p')}\right|^\half_k
  \left| \frac{\partial(\b_{12},\b_3)}{\partial(\P, \p)}\right|^\half_k
  \\ \nonumber
  &&\quad\times
\phi_\lambda^*(\k')
  \< \b_{12}', \b_3', \k'  | j(0) | \b_{12}, \b_3, \k \>
  \phi_\lambda(\k) {\sqrt{\omega_m (\mathbf{p}_3)}} .
\end{eqnarray}
The suffix $k$ in the Jacobian indicates that the non-interacting
two-body invariant mass, $M_{k} $(Eq.~\ref{eq:b4c}), was used in the
transformation of Eq.~(\ref{eq:b7a}), while the suffix $\lambda$
indicates that the mass of the bound pair, $M_\lambda$, was used to
compute the variable change.  The $\mathbf{k}$-dependence of Jacobians
with the subscript $k$ is limited to the magnitude of $\mathbf{k}$.

Since the scalar density operator $j(x)$ operates only in the space of
particle~3, we have
\begin{equation}
  \label{eq:b12}
\sqrt{\omega_m (\mathbf{b}_3')}
  \< \b_{12}', \b_3', \k'  | j(0) | \b_{12}, \b_3, \k \>
\sqrt{\omega_m (\mathbf{b}_3')}
  = m \delta(\k' - \k) \delta(\b_{12}' - \b_{12}) G[(b_3' - b_3)^2].
\end{equation}
The integral over $\p_{12}'$ can be converted to an integral over
$\b_{12}'$. The Jacobian of the variable change is
\begin{equation}
  \label{eq:b13}
  \int d\p_{12}' = \int d\b_{12}' 
  \left| \frac{\partial(\P', \p')}{\partial(\b_{12},'\b_3')}\right|_{k}
  \left| \frac{\partial(\p_{12}',\p_3')}{\partial(\P', \p')}\right|_\lambda.
\end{equation}
The final result is
\begin{equation}
  \label{eq:b14}
    \F_{BT}^{\rm instant} = 
    \int d\k {\sqrt{\omega_m (\mathbf{p}'_3)} \over 
\sqrt{\omega_m (\mathbf{b}'_3)}}
\left| \frac{\partial(\p_{12}',\p_3')}{\partial(\P', \p')}\right|^\half_\lambda
  \left| \frac{\partial(\P, \p)}{\partial(\p_{12},\p_3)}\right|^\half_\lambda
  \left| \frac{\partial(\P', \p')}{\partial(\b_{12}',\b_3')}\right|^\half_{k}
  \left| \frac{\partial(\b_{12},\b_3)}{\partial(\P, \p)}\right|^\half_{k}
  \left|\phi_\lambda(\k)\right|^2 
{\sqrt{\omega_m (\mathbf{p}_3)}
\over \sqrt{\omega_m (\mathbf{b}_3)}}
m  G[(b_3' - b_3)^2].
\end{equation}

The relevant observation is that this integral has a non-trivial
dependence on the momentum, $\p_{12}$, of the bound state, in contrast
to the $\p_{12}$ independence of the tensor product result.  In this
model the $BT$ and tensor product representations of the
three-particle system are related by one of Sokolov's unitary
transformations, $A$, that preserve the three-body $S$-matrix:
\begin{equation}
| \lambda , m; \p_{12}, \p_3 \>_{TP} = A | \lambda , m; \p_{12}, \p_3 \>_{BT}.  
\label{eq:b14.a}
\end{equation}
The scale of the $\p_{12}$ dependence in Eq.~(\ref{eq:b14}) provides a
measure of the size of the violation of cluster properties that
results from ignoring $A$ by replacing it with the identity.  
Since we know both states in Eq.~(~\ref{eq:b14.a}) we do not
have to calculate $A$ explicitly to determine its impact.

For this calculation, we vary the three-momentum transfer 
$\q$ and the spectator momentum $\p_{12}$.  The initial
momentum of particle 3 and the final momentum of the 
system are fixed in terms of these variables:
\begin{itemize}
\item $\p_3 = -\half\q$;
\item $\P' = \P + \q$.
\end{itemize}
Since in the instant form,
\begin{equation}
  \label{eq:b15}
  \P = \B; \qquad
  \P' = \B',
\end{equation}
and the scalar density matrix element constrains $\b_{12}' = \b_{12}$, we
find that $\b_3' = \b_3 + \q$.  However, the final momenta
$\p_{12}'$ and $\p_3'$ are not constrained, and in general there are
non-vanishing contributions to this matrix element for $\p_3' \ne
\p_3 + \q$ and $\p_{12}' \ne \p_{12}$.

In the nonrelativistic limit, where $\k$, $\q$ and $\p_{12}$ are all
small with respect to the relevant masses, the Jacobians are
approximately unity and can be factored out of the integral, leaving a
unit wave function normalization and a result identical to the $TP$
case.  The quantitative level of disagreement with the $TP$ result is
therefore linked to the extent to which the model goes beyond the
nonrelativistic limit.

For the other forms of dynamics we preform similar calculations.  
However, rather than using the natural variables for each given form of
dynamics (four-velocity, light-front components of the four-momentum),
we use the same variables employed in the instant-form case to
facilitate a comparison of the size of the violations of cluster
properties. These variables are also the quantities that are most
readily measured using detectors.  Calculations using front and
point-form dynamics lead to results that have the same general form as
Eq.~(\ref{eq:b14}) involving an integral over Jacobians and their
inverses involving different sets of variables and the bound state
wave functions.  The Jacobians all cancel in the limit that
two-body mass eigenvalue becomes the two-body invariant mass.

\subsection{$BT$ Representation: Front Form}

Dirac's front-form dynamics is described in detail in
Ref.~\cite{Keister:1991sb}.  In a front-form dynamics the generators
of transformations that leave a plane tangent to the light cone
invariant are free of interactions.  We provide a summary here.

Basis states in the front form are eigenstates of the light-front components 
of the momenta 
\begin{equation}
  \label{eq:f1}
  \fp = (\p_\perp,p^+);\quad p^+ = p^0 + p^3; \quad \p_{\perp}=(p^1,p^2).
\end{equation}
These operators generate translations in a hyperplane $x^-:= x^0-x^3
=0$ tangent to the light cone.  In what follows we use a tilde to
indicate the light-front components of four-vectors.

In the front form the Lorentz transformations Eq.~(\ref{eq:b2}) and
Eq.~(\ref{eq:b7a}) used to define a particle momentum in the rest
frame of the three-particle system in the instant form are replaced by
boosts that leave the light front invariant.

For the front-form Bakamjian-Thomas construction Eq.~(\ref{eq:b5}) is
replaced by 
\beq
\tilde{\B} = \tilde{\P}; \quad \b = \p.
\label{eq:b5d}
\eeq
As in the instant-form case the first equation is a general property
of a light-front dynamics while the second equation is an additional
requirement of the Bakamjian-Thomas construction.  The second equation
identifies the momentum of particle~3 in the rest frame of the
non-interacting three-particle system, $\p$, with the momentum, $\b$
of particle~3 in the rest frame of the $(12) +3$ system.  As in the
instant-form case this identification has no consequences for the
two-body $S$ matrix, but will lead to a violation of cluster
separability at the four-body level.  With this modification the
front-form BT expression for $\F_{BT}^{\rm front}$ is similar to the
instant-form result:
\begin{equation}
  \label{eq:f2}
    \F_{BT}^{\rm front} := \int d\k
\sqrt{{{p}^{+\prime}_3
\over {b}^{+\prime }_3}} 
\left| \frac{\partial(\fp_{12}',\fp_3')}
{\partial(\fP', \p')}\right|^\half_\lambda
    \left| \frac{\partial(\fP, \p)}
{\partial(\fp_{12},\fp_3)}\right|^\half_\lambda
        \left| 
\frac{\partial(\fP', \p')}{\partial(\fb_{12}',\fb_3')}\right|^\half_{k}
    \left| 
\frac{\partial(\fb_{12},\fb_3)}{\partial(\fP, \p)}\right|^\half_{k}
    \left|\phi_\lambda(\k)\right|^2
\sqrt{{p}^{+}_3
\over {b}^+_3}
m G[(b_3' - b_3)^2].
\end{equation}
This quantity has an unphysical dependence on $\fb_{12}$ that does
not occur in the tensor product. 

The instant- and front-form calculations differ in both the constraints
(\ref{eq:b5d}) vs (\ref{eq:b5d}) and the different boosts use to
construct $\p, \p',\b$ and $\b'$. For the purpose of the
comparison with the instant-form calculations we make a final variable
change replacing the light-front components of the momenta by the same
variables that were used in the instant-form calculation.

For this calculation, we vary $\fq$ and $\fp_{12}$, with coordinate
axes chosen such that $q^+ = 0$ and $p_{12}^+ = 0$.  We define
$\p_\perp$ and $\fP'$ in terms of these quantities:
\begin{itemize}
\item $\p_\perp = -\half\q_\perp$;
\item $\fP' = \fP + \fq$.
\end{itemize}
Analogous to the discussion following Eq.~(\ref{eq:b15}), we have that 
\begin{equation}
  \label{eq:f3}
  \fP = \fB \qquad
  \fP' = \fB',
\end{equation}
The scalar density matrix element constrains $\fb_{12}' = \fb_{12}$, and
therefore $\fb_3' = \fb_3 + \fq$.  However, the final
momenta $\fp_{12}'$ and $\fp_3'$ are not constrained, and in general
the integral has non-zero contributions from
$\fp_3' \ne \fp_3 + \fq$ and $\fp_{12}' \ne \fp_{12}$.

The constraints $\p_\perp = -\half\q_\perp$;
and $\fP' = \fP + \fq$ ensure that the momentum transfer to the 
one-body system is q, as in the instant form example.   

\subsection{$BT$ Representation: Point Form}

Dirac's point-form dynamics are also described in detail in
Ref.~\cite{Keister:1991sb}.  In this case the Lorentz group
is non-interacting.  We provide a summary here.

Basis states in the point form are described by three-velocity vectors
$\v$.  Momenta are obtained by multiplying the four-velocities by
(interacting or non-interacting) masses.  Thus, we seek to evaluate
matrix elements of the scalar density operator $j(x)$ between states of
particle~3 and the (12) bound state with initial three-velocity $\V=\P/M$
and final three-velocity $\V'=\P'/M'$.

In the point-form Bakamjian-Thomas construction, equations
(\ref{eq:b5}) and \ref{eq:b5d} are replaced by
\beq
\V = \B/M_0; \quad \b = \p.
\label{eq:b5e}
\eeq
The first of these equations corresponds to total four-velocity
conservation, which is a property of all point-form representations.
The boosts in the point-form are the same as those used in the instant
form, but they are parameterized by the conserved four-velocities.
The second equation, as in the instant and from form, is a requirement
of the Bakamjian-Thomas construction.

The derivations proceed in a fashion similar to the instant form.

With these conventions, the point-form 
$\F_{BT}$ has the structure
\begin{equation}
  \label{eq:p1}
    \F_{BT}^{\rm point} := \int d\k
\sqrt{{\omega_m (\mathbf{p}'_3)
\over \omega_m (\mathbf{b}'_3)}}
    \left| \frac{\partial(\p_{12}',\p_3')}{\partial(\V' ,\p')}\right|^\half_\lambda
    \left| \frac{\partial(\V, \p)}{\partial(\p_{12},\p_3)}\right|^\half_\lambda 
    \left| \frac{\partial(\V', \p')}{\partial(\b_{12}',\b_3')}\right|^\half_k
    \left| \frac{\partial(\b_{12},\b_3)}{\partial(\V, \p)}\right|^\half_k
    \left|\phi_\lambda(\k)\right|^2
\sqrt{{\omega_m (\mathbf{p}_3)
\over \omega_m (\mathbf{b}_3)}}
 m G[(b_3' - b_3)^2].
\end{equation}


Note in this case that the velocities $\V$ and $\V'$ are the same for
the kinematics using the interacting mass $M$ or the free mass
$M_{0}$.  Since the boosts depend on $\V$ it follows that
\begin{equation}
  \label{eq:p2}
  \b_3 = \p_3;\quad \b_3' = \p_3',
\end{equation}
and therefore that $\b_3' -\b_3 = \p_3'-\p_3 = \q$.  One consequence
is that in general, $M'\V'-M \V \ne \q$.

We constrain the momentum transfer to particle 3:
\begin{itemize}
\item $\b_3 = -\half\q$;
\item $\b_3' = \b_3 + \q$.
\end{itemize} 

These calculations share the property that $\b_3'-\b_3 = \q$ in the
instant and point forms, and $\tilde\b_3' - \tilde\b_3 = \tilde\q$ in
the front form, i.e.\ the constituent-particle momenta are related to
each other at the three-vector level by the momentum transfer to the
overall three-body system.  One could choose different sets of
particle momenta in calculating $BT$ matrix elements to compare to the
$TP$ result.  For example, we could constrain the three calculations
such that $P'-P =q$ in some frame.  Different choices explore
different kinematic regions of the difference between $A$ and the
identity.  Our investigations of other choices indicate numerical
effects of comparable size.

\section{Results and Discussion}

In this section we discuss results of calculations of the form factor
$\F_{BT}$ of a scalar density for instant-, front- and  point-form
kinematic choices.  Initially we consider scales that are relevant for
systems of nucleons interacting with two-body interactions.  Then we
turn to examples more appropriate to hadronic models with subnucleon
degrees of freedom.

In all figures we display the figure of merit:
\begin{equation}
  \label{eq:p10}
  \frac{(\F_{BT}-\F_{TP})}{\F_{TP}},
\end{equation}
which represents the relative error induced by ignoring the unitary
transformation Eq.~(\ref{eq:b14.a}) that restores cluster
separability.

\subsection{Malfliet-Tjon Deuteron Wave Function}

To model realistic conditions for nuclear physics, we use a
bound-state spectator with deuteron properties constructed from
Malfliet-Tjon~\cite{Malfliet:1968tj} potential IV, which contains both
attractive and repulsive forces.  Ref.~\cite{Keister:1991sb} discusses
how to make a phase-equivalent relativistic model that has the same
wave functions as the non-relativistic model, and our numerical
results follow that approach.  The invariant form factor $G(q^2)$
is taken as a dipole form factor.

We first consider the figure of merit as a function of the momentum
transfer and momentum of the bound state in each of Dirac's forms of
dynamics and for the bound state momentum perpendicular and parallel
to the momentum transfer.  The results are shown in figures 2-7.
In all cases the expected results are given by the flat
planes.

The fractional deviation of this $BT$ model calculation from the $TP$
result that satisfies cluster separability is very small for 
all three forms of dynamics,  typically of order $10^{-3}$ or smaller
at the highest values of the three-momenta $\q$ and $\p_{12}$.   

\begin{figure*}[h]
  \begin{minipage}{0.45\linewidth} \centering
    \includegraphics[width=\linewidth]{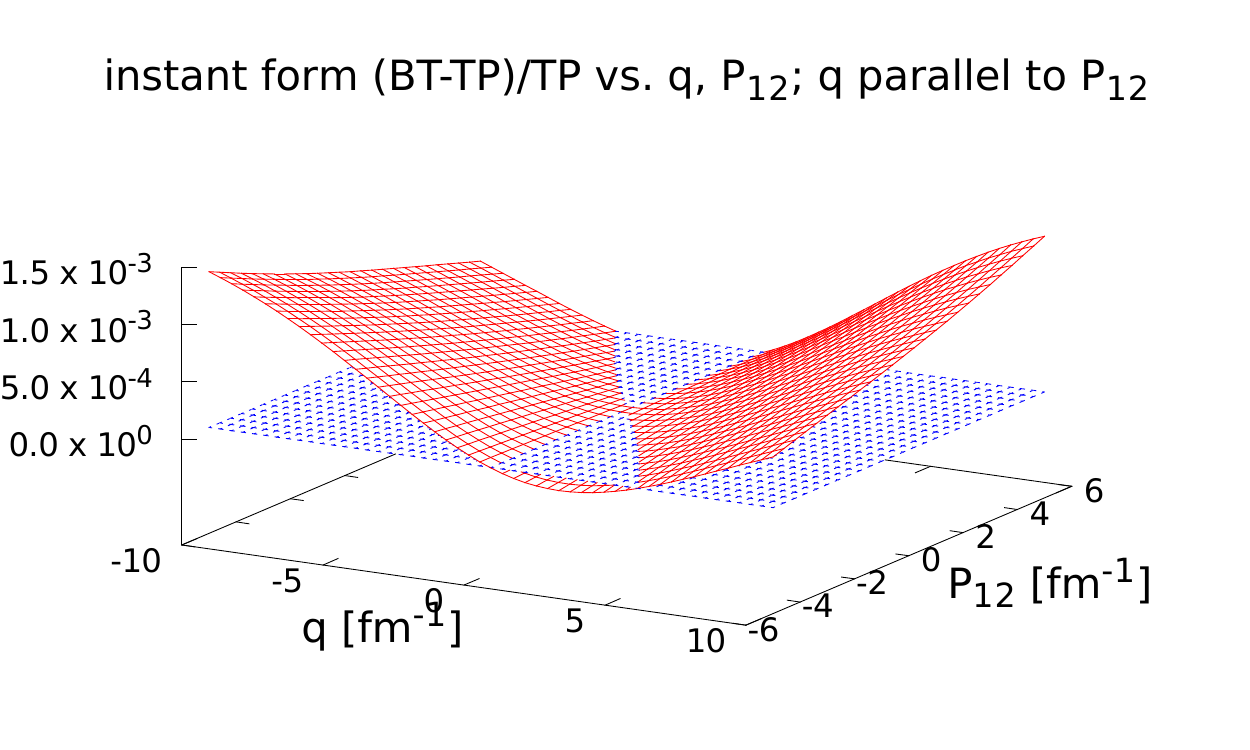}
    \caption{(Color online) Model differences for instant-form $BT$ calculation,
$\q\|\p_{12}$.}
    \label{fig:instant-par}
  \end{minipage} \hspace{0.05\linewidth}
  \begin{minipage}{0.45\linewidth} \centering
    \includegraphics[width=\linewidth]{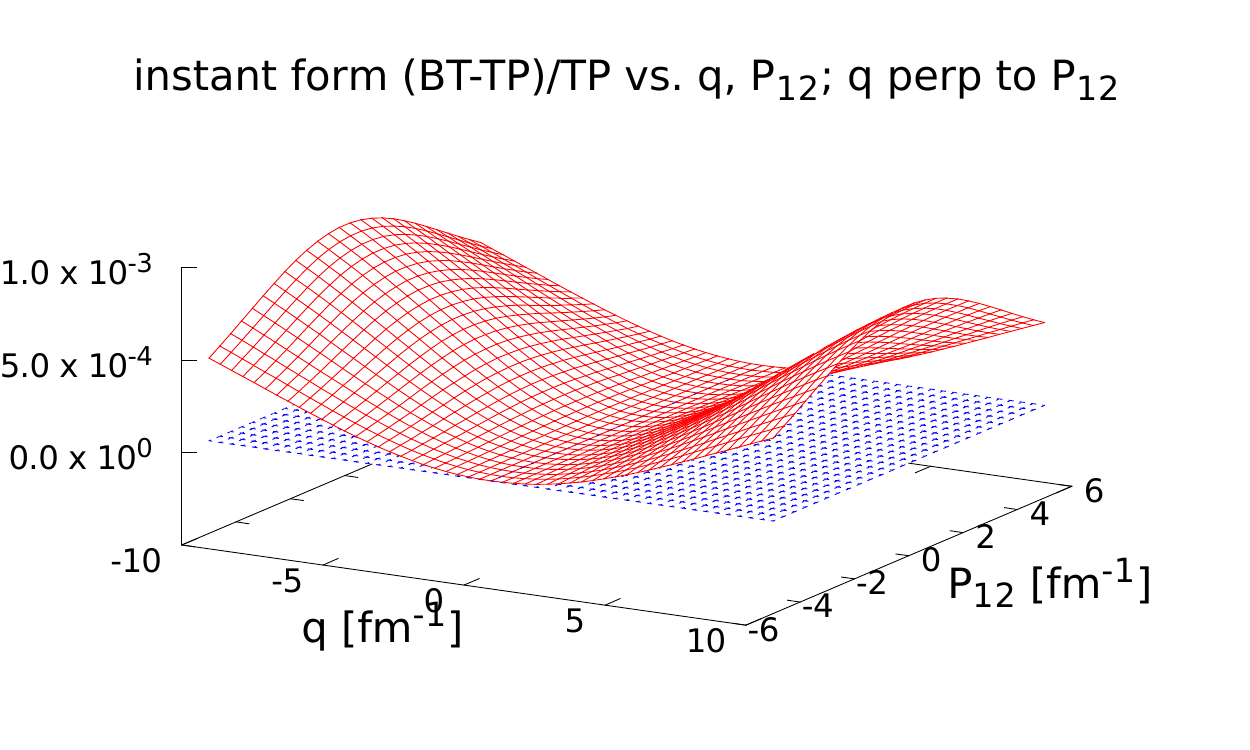}
    \caption{(Color online) Model differences for instant-form $BT$ calculation,
$\q\perp\p_{12}$.}
    \label{fig:instant-prp}
  \end{minipage}
\end{figure*}

\begin{figure*}[h]
  \begin{minipage}{0.45\linewidth}
    \centering
    \includegraphics[width=\linewidth]{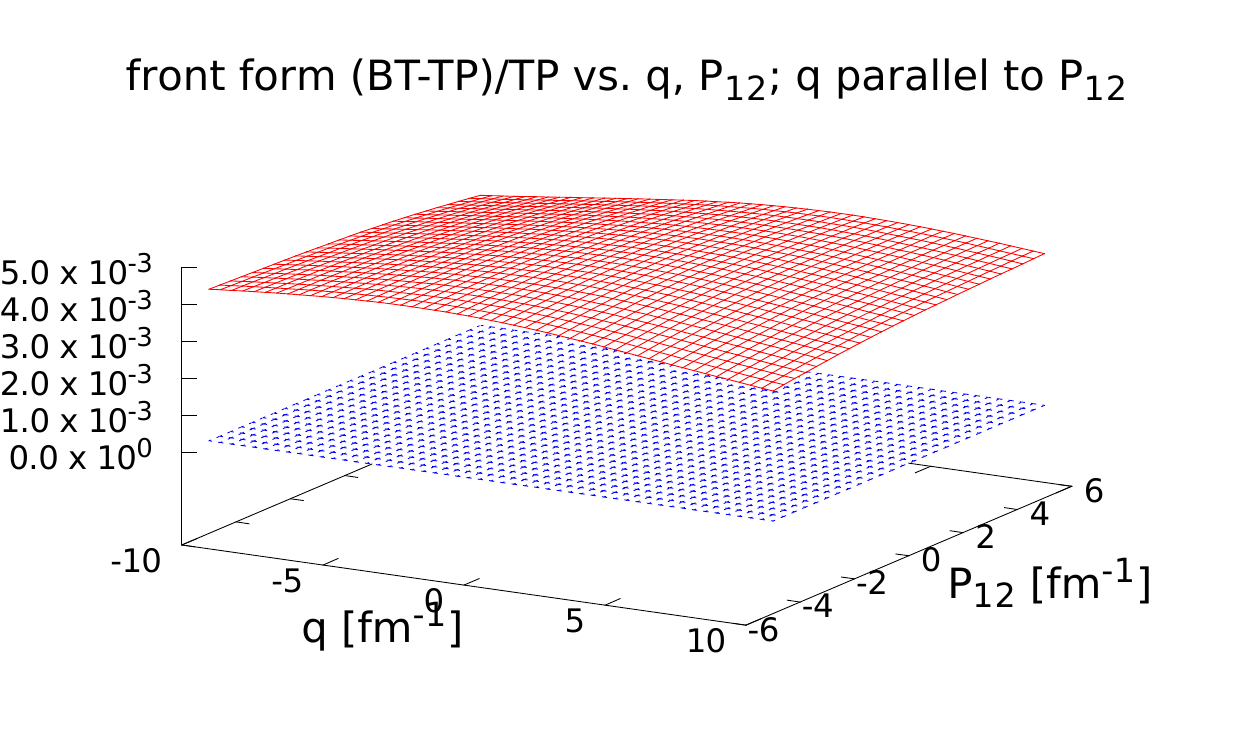}
    \caption{(Color online) Model differences for front-form $BT$ calculation, $\q\|\p_{12}$.}
    \label{fig:front-par}
  \end{minipage}
  \hspace{0.05\linewidth}
  \begin{minipage}{0.45\linewidth}
    \centering
    \includegraphics[width=\linewidth]{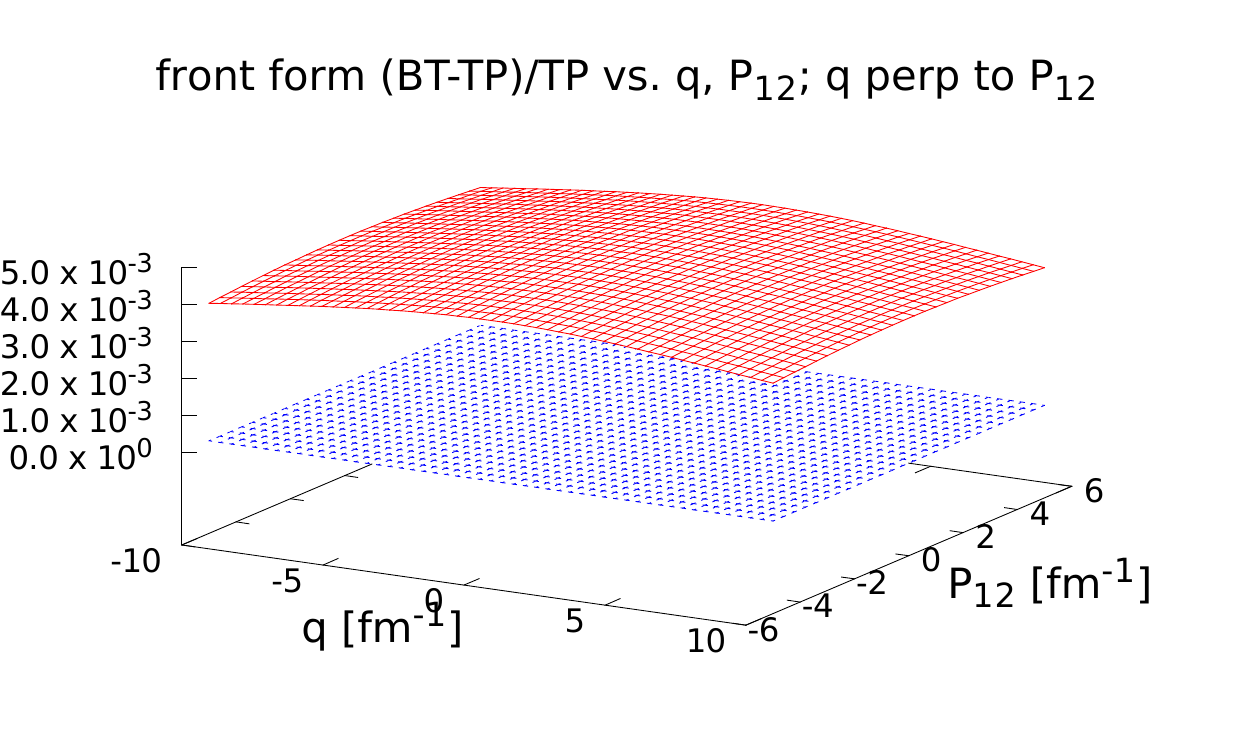}
    \caption{(Color online) Model differences for front-form $BT$ calculation, $\q\perp\p_{12}$.}
    \label{fig:front-prp}
  \end{minipage}
\end{figure*}

\begin{figure*}[h]
  \begin{minipage}{0.45\linewidth}
    \centering
    \includegraphics[width=\linewidth]{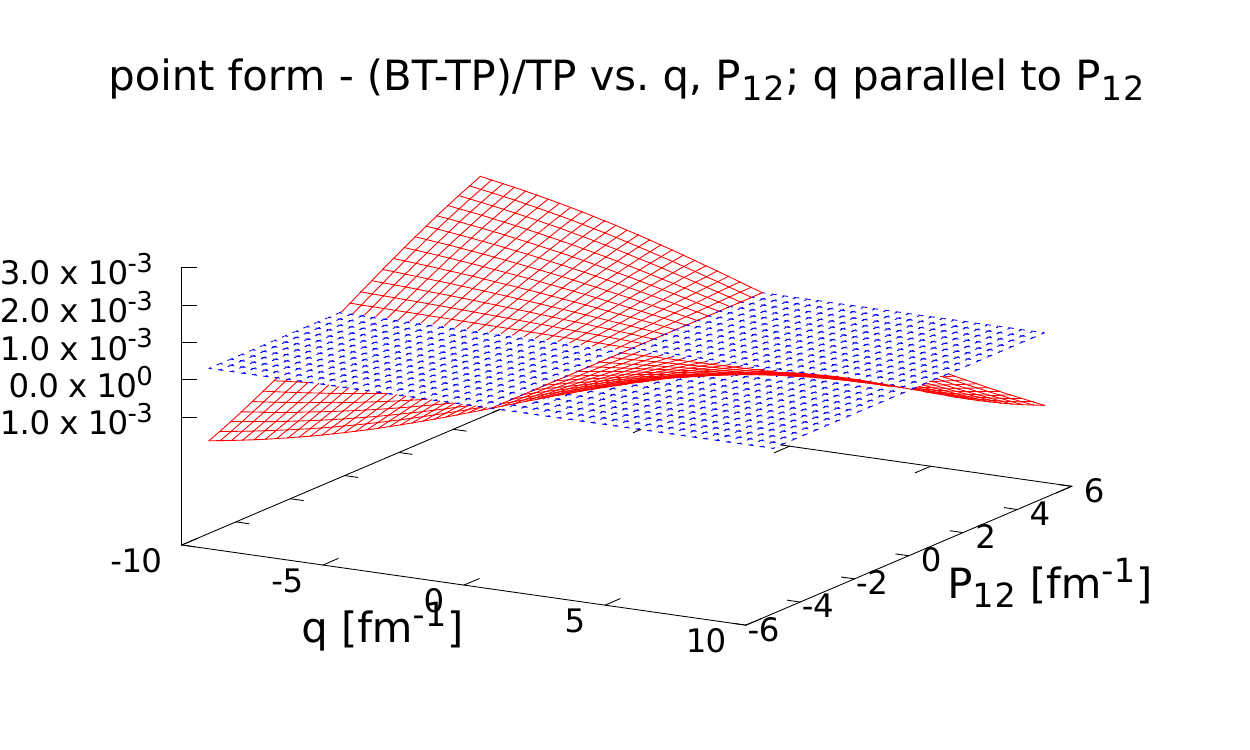}
    \caption{(Color online) Model differences for point-form $BT$ calculation A, $\q\|\p_{12}$.}
    \label{fig:point-par}
  \end{minipage}
  \hspace{0.05\linewidth}
  \begin{minipage}{0.45\linewidth}
    \centering
    \includegraphics[width=\linewidth]{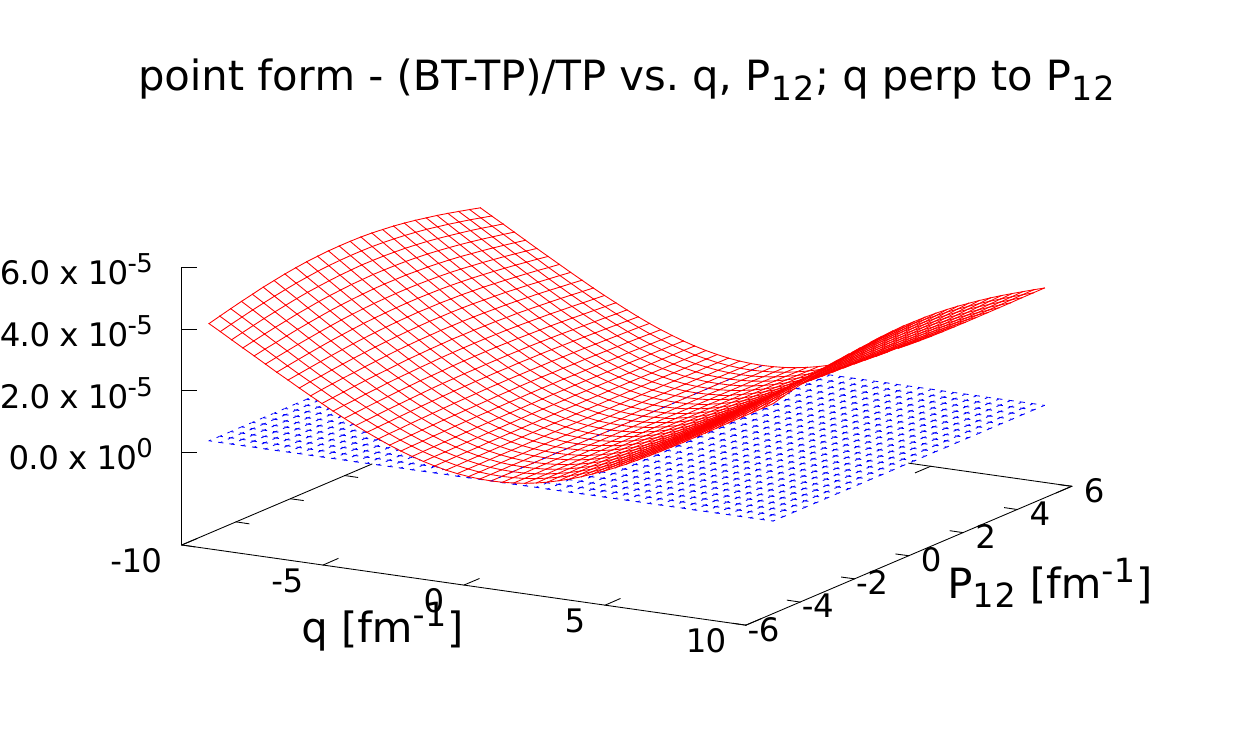}
    \caption{(Color online) Model differences for point-form $BT$ calculation A, $\q\perp\p_{12}$.}
    \label{fig:point-prp}
  \end{minipage}
\end{figure*}


\subsection{Binding Energy Variation}

\begin{figure*}[h]
  \begin{minipage}{0.45\linewidth}
    \centering
   \includegraphics[width=\linewidth]{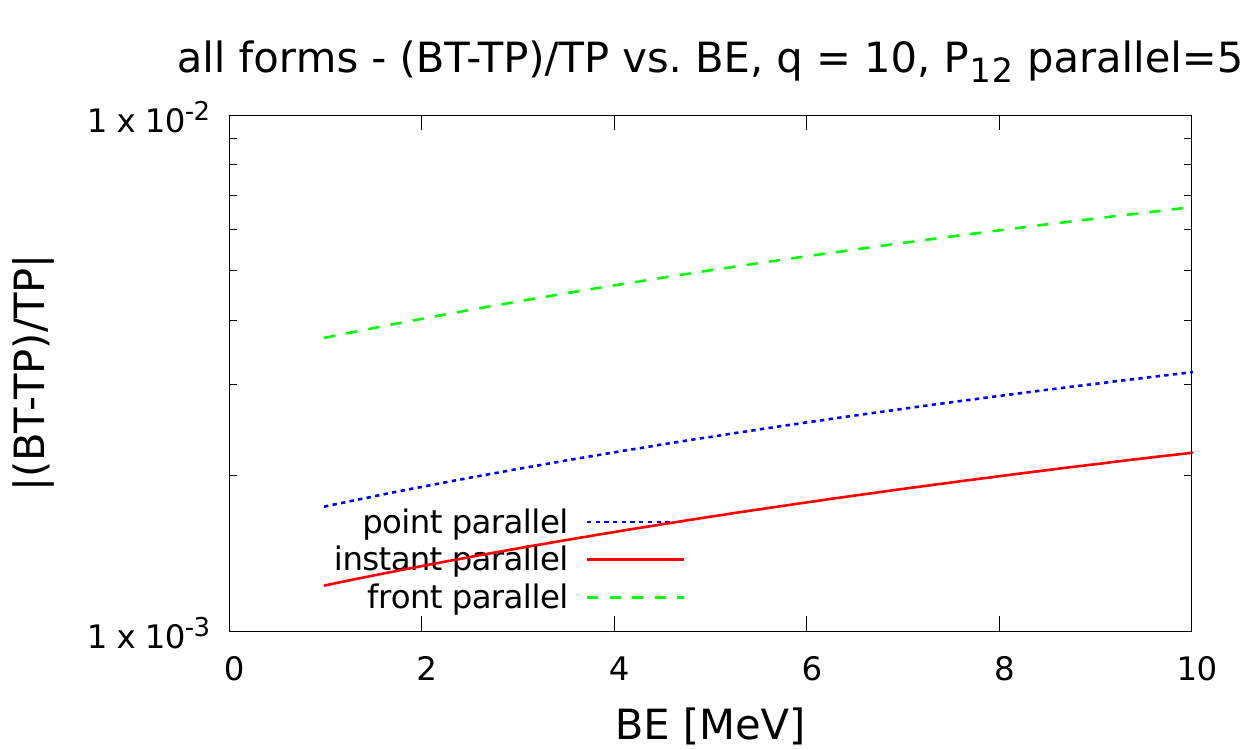}
    \caption{(Color online) Model differences for $BT$ calculations as a function of
      two-body spectator binding energy, $\q=10~{\rm fm}^{-1}\|\p_{12}$.}
    \label{fig:MT-BE-par}
  \end{minipage}
  \hspace{0.05\linewidth}
  \begin{minipage}{0.45\linewidth}
    \centering
    \includegraphics[width=\linewidth]{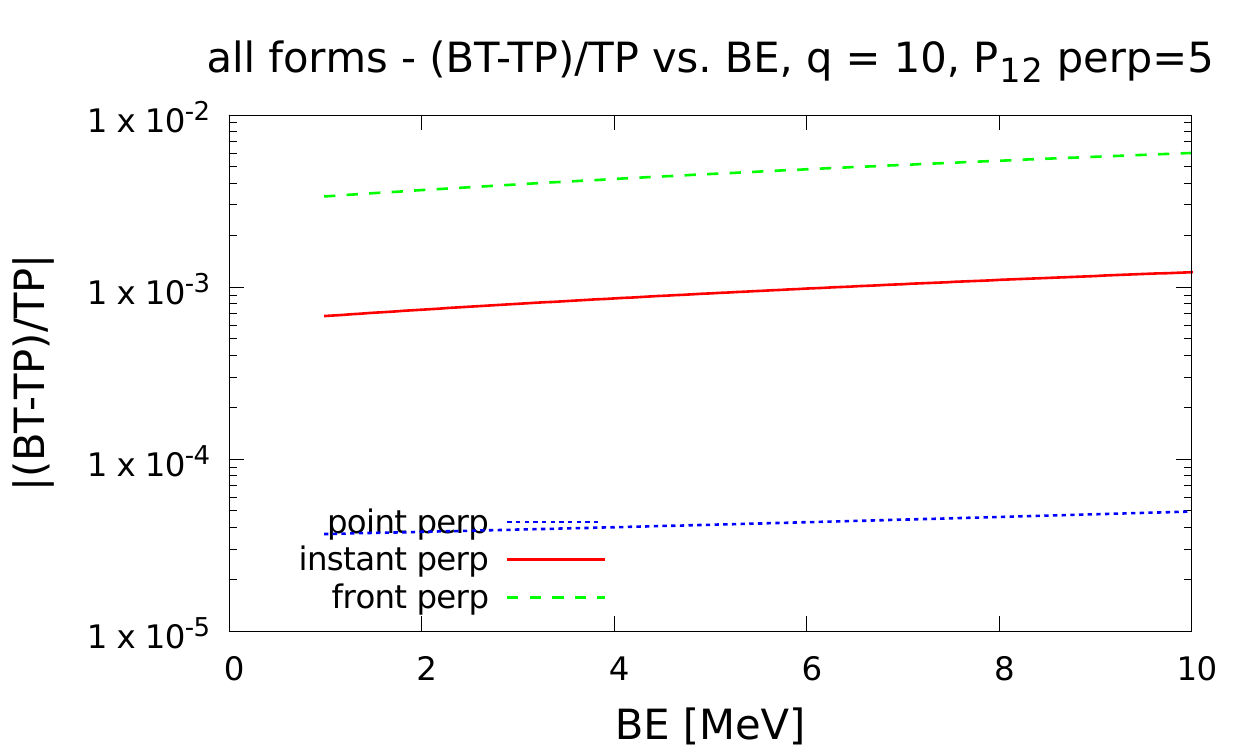}
    \caption{(Color online) Model differences for $BT$ calculations as a function of
      two-body spectator binding energy, $\q=10~{\rm fm}^{-1}\perp\p_{12}$.}
    \label{fig:MT-BE-perp}
  \end{minipage}
\end{figure*}

The calculations above assumed a bound state with a wave function
having a typical dependence on the relative momentum of the
constituent nucleons.  The next set of curves illustrates the figure
of merit for fixed values of $\q$ and $\p_{12}$ as we vary the binding
energy and momentum scale of the wave function.  The variations that
we consider are still scales that are relevant to nuclear systems.
Figures~\ref{fig:MT-BE-par} and~\ref{fig:MT-BE-perp} show the results
of calculations for varying binding energy with a Malfliet-Tjon wave
function.

As with the earlier cases that employed the bound-state binding energy,
the fractional deviation of the $BT$ results from the $TP$ benchmark is
quite small, of order $10^{-3}$ or less.

\subsection{Wave Function Scale Variation}

\begin{figure*}[h]
  \begin{minipage}{0.45\linewidth}
    \centering
    \includegraphics[width=\linewidth]{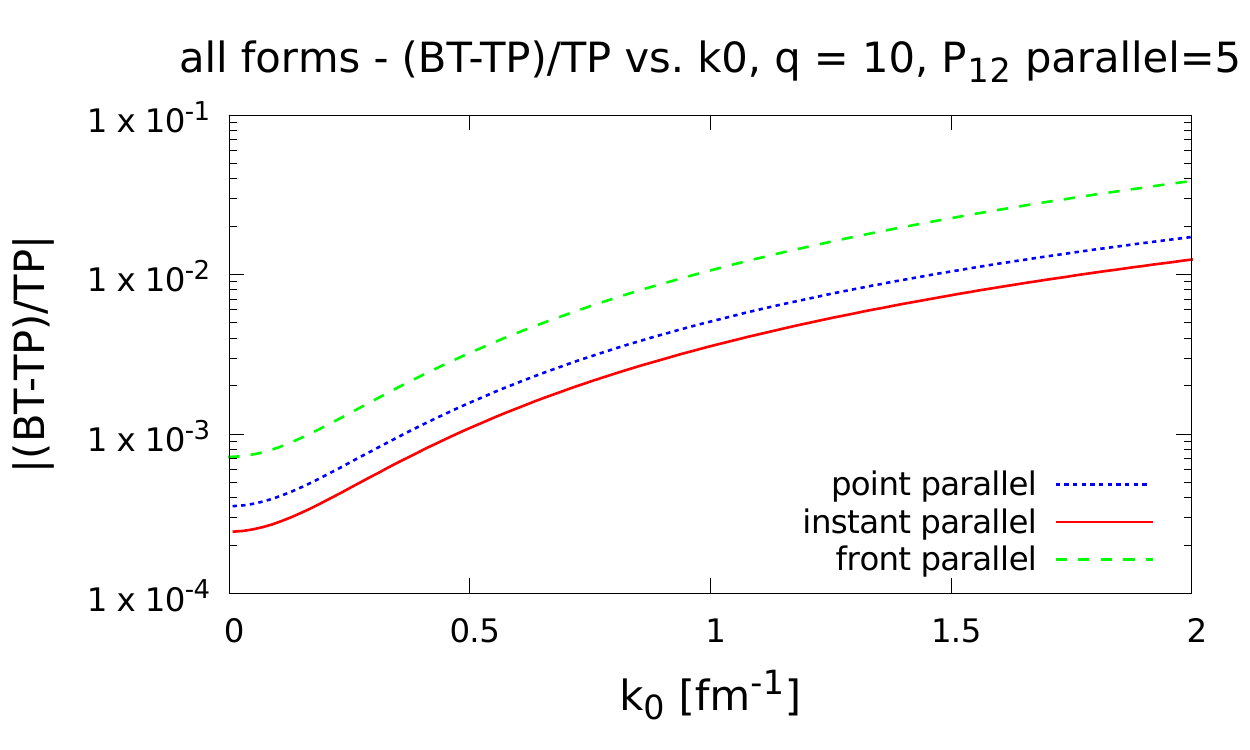}
    \caption{(Color online) Model differences for $BT$ calculations as a function of
      two-body Gaussian wave function scale, $\q=10~{\rm fm}^{-1}\|\p_{12}$.}
    \label{fig:Gaussian-par}
  \end{minipage}
  \hspace{0.05\linewidth}
  \begin{minipage}{0.45\linewidth}
    \centering
    \includegraphics[width=\linewidth]{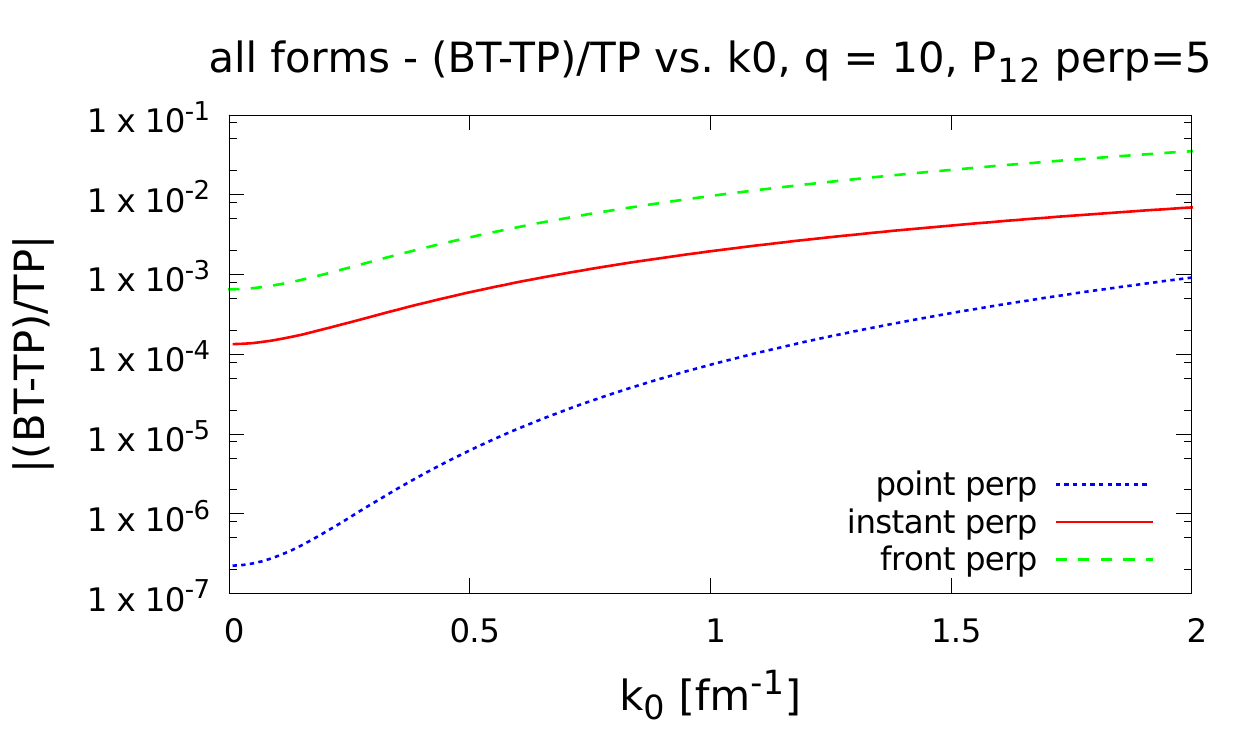}
    \caption{(Color online) Model differences for $BT$ calculations as a function of
      two-body Gaussian wave function scale, $\q=10~{\rm fm}^{-1}\perp\p_{12}$.}
    \label{fig:Gaussian-perp}
  \end{minipage}
\end{figure*}

We also examined sensitivity to the momentum scale, $k_0$, of the wave
function by replacing the Malfliet-Tjon wave function with a Gaussian
form:
\begin{equation}
  \label{eq:g1}
  \phi(\k) = {1\over\sqrt{N}} e^{-(k/k_0)^2}.
\end{equation}
Figures~\ref{fig:Gaussian-par} and~\ref{fig:Gaussian-perp} show
the results for the spectator momentum perpendicular and parallel to
the momentum transfer in all three forms of dynamics.

These results mirror those discussed above: the fractional deviation
of the $BT$ results from the $TP$ benchmark is quite small, of order
$10^{-3}$ or less.

\subsection{Implications for Nuclear Theory}

Formally dynamical relativistic models should be Poincar\'e invariant
and satisfy cluster properties; in addition operators associated with
short ranged physical phenomena should have well defined cluster
expansions.


While our model is considerably simpler than a realistic model of a
three-nucleon system interacting with an electron, the difference
between the Bakamjian-Thomas formulation of the model and the model
that clusters properly is due entirely to the same unitary
transformations discussed in our simplified model.  This difference
vanishes in the limit that these transformations become the identity.
The analysis in this paper established that these operators, which
implicitly depend on the interactions, are close to the identity for
interactions with typical nuclear physics scales for binding energy
and Fermi momentum.  This provides a strong justification for applying
Bakamjian-Thomas construction of dynamical representations Poincar\'e
group to system of more than three nucleons.

Our results indicate that Bakamjian-Thomas models, which explicitly
satisfy the requirements of Poincar\'e invariance, can be utilized in
typical nuclear physics problems with minimal quantitative error due
to the lack of cluster separability using any of Dirac's front-form
dynamics.  

\subsection{Implications for Hadron Models}

The final set of figures show the results of calculations with scales
that are more appropriate models of hadrons based on sub-nucleonic
degrees of freedom.  

We note here that confinement precludes separating arbitrary
subsystems by large distance scales, so the requirement of cluster
separability is irrelevant for models of individual hadrons.  The
issue may be relevant, however, for systems of hadrons described by
sub-nucleonic degrees of freedom.

In these cases, in order to understand the relevant scales, we replace
the particle~3 mass in the above calculations by a ''constituent quark''
mass of 220 MeV, and consider two-body (``diquark'') masses ranging
from 200 to 600 MeV for Gaussian wave functions with a 1~fm${}^{-1}$
scale in Figs. 12 and 13, and Gaussian wave functions with scale ranging from
0.5~fm${}^{-1}$ to 10~fm${}^{-1}$, with a diquark mass of 600 MeV in Figs. 
14 and 15.

The results are similar to those appropriate to nuclear physics
discussed above, except that the scale of deviation from the $TP$
benchmark is on the scale of 10\% for the instant, front and
point-form $A$s.  This not surprising given that the mass/momentum
scale variation for these calculations is much higher than for typical
cases in nuclear physics with nucleons.  While these corrections may
be relevant, multi-hadron models based on sub-nucleon degrees of
freedom have not been developed to a precision where these effects
could be identified.

\begin{figure*}[h]
  \begin{minipage}{0.45\linewidth}
    \centering
    \includegraphics[width=\linewidth]{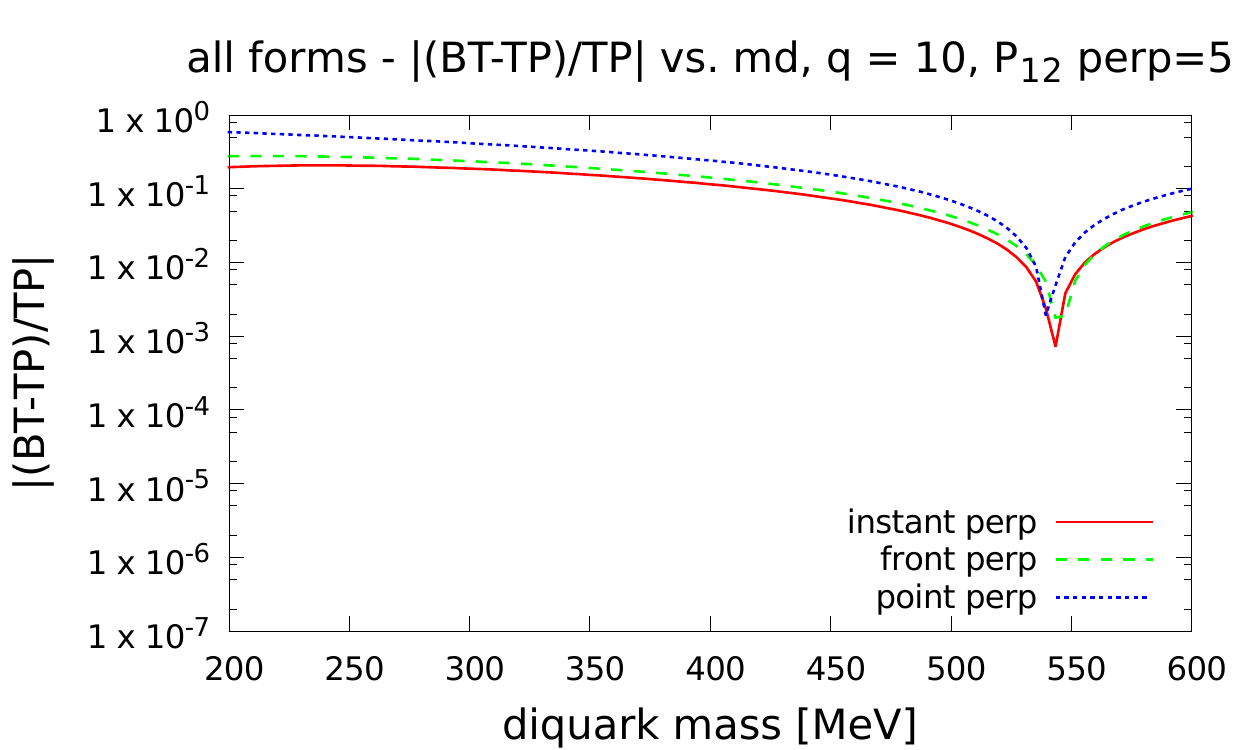}
    \caption{(Color online) Model differences for $BT$ calculations as a function of
      two-body spectator mass, $\q=10~{\rm fm}^{-1}\|\p_{12}$.}
    \label{fig:quark-BE-all-par}
  \end{minipage}
  \hspace{0.05\linewidth}
  \begin{minipage}{0.45\linewidth}
    \centering
    \includegraphics[width=\linewidth]{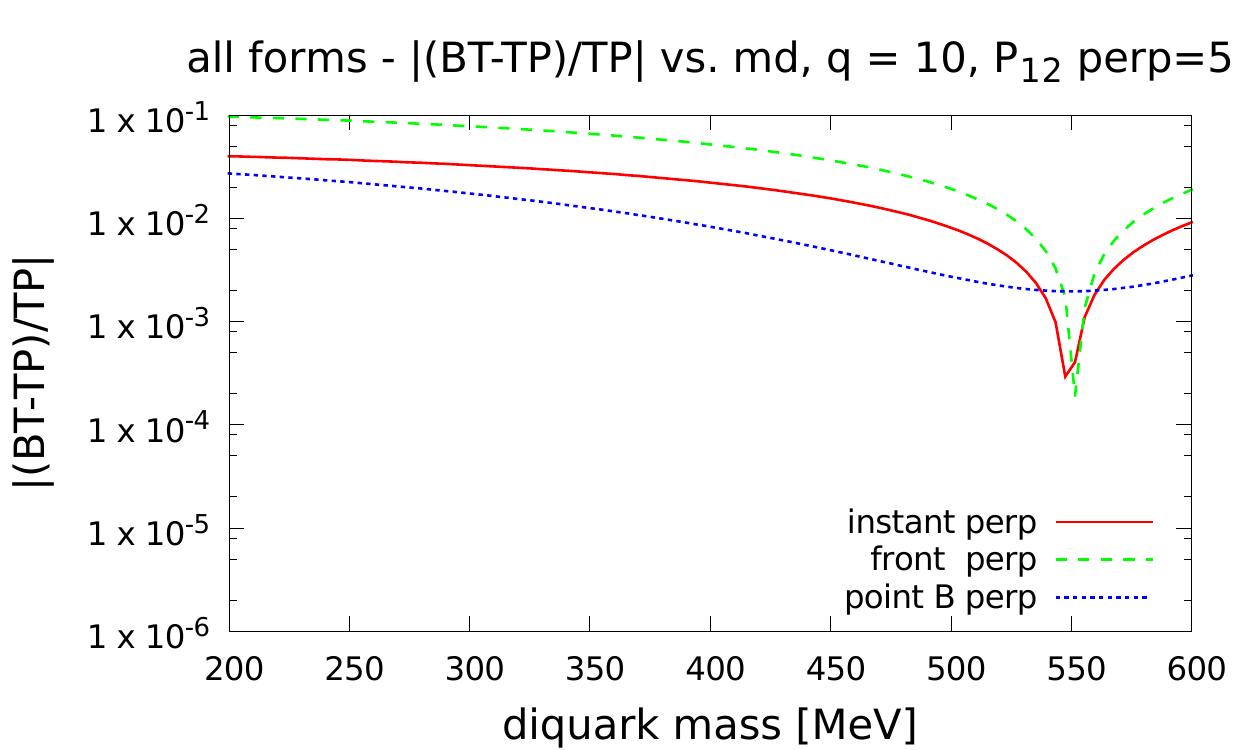}
    \caption{(Color online) Model differences for $BT$ calculations as a function of
      two-body spectator mass, $\q=10~{\rm fm}^{-1}\perp\p_{12}$.}
    \label{fig:quark-BE-all-prp}
  \end{minipage}
\end{figure*}

\begin{figure*}[h]
  \begin{minipage}{0.45\linewidth}
    \centering
    \includegraphics[width=\linewidth]{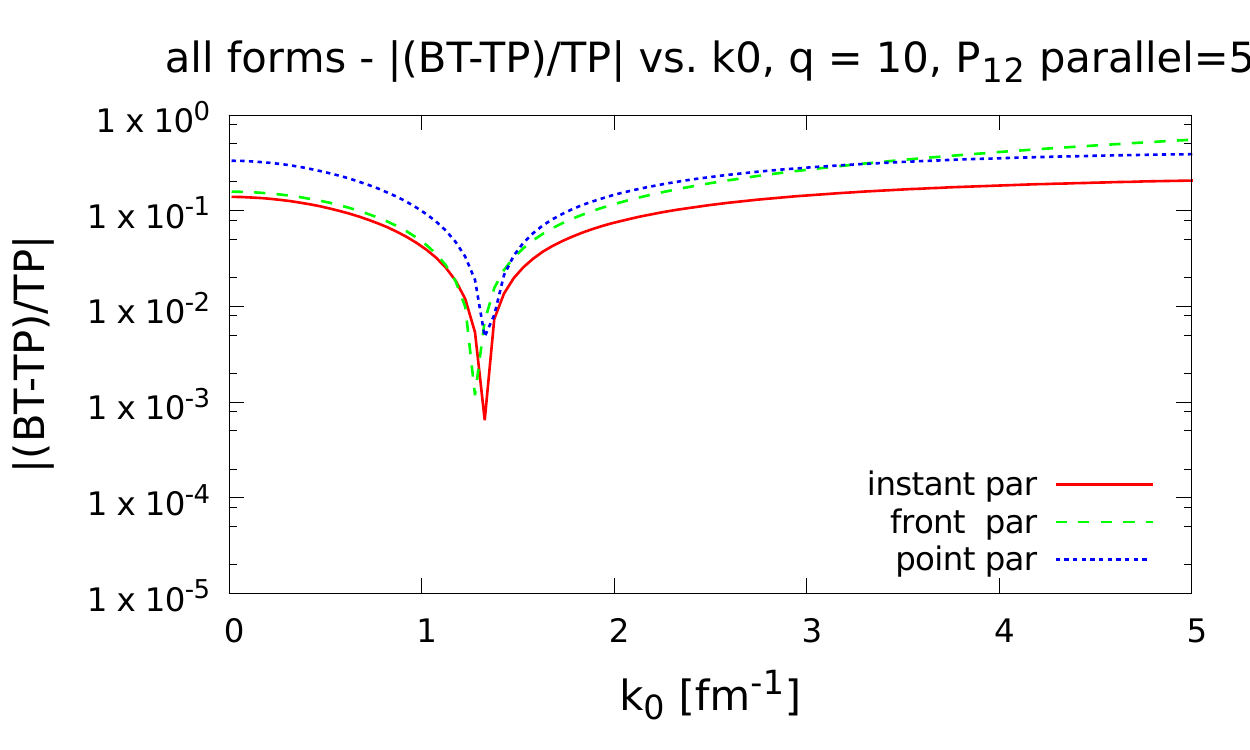}
    \caption{(Color online) Model differences for $BT$ calculations as a function of
      Gaussian wave function scale, $\q=10~{\rm fm}^{-1}\|\p_{12}$.}
    \label{fig:quark-k0-all-par}
  \end{minipage}
  \hspace{0.05\linewidth}
  \begin{minipage}{0.45\linewidth}
    \centering
    \includegraphics[width=\linewidth]{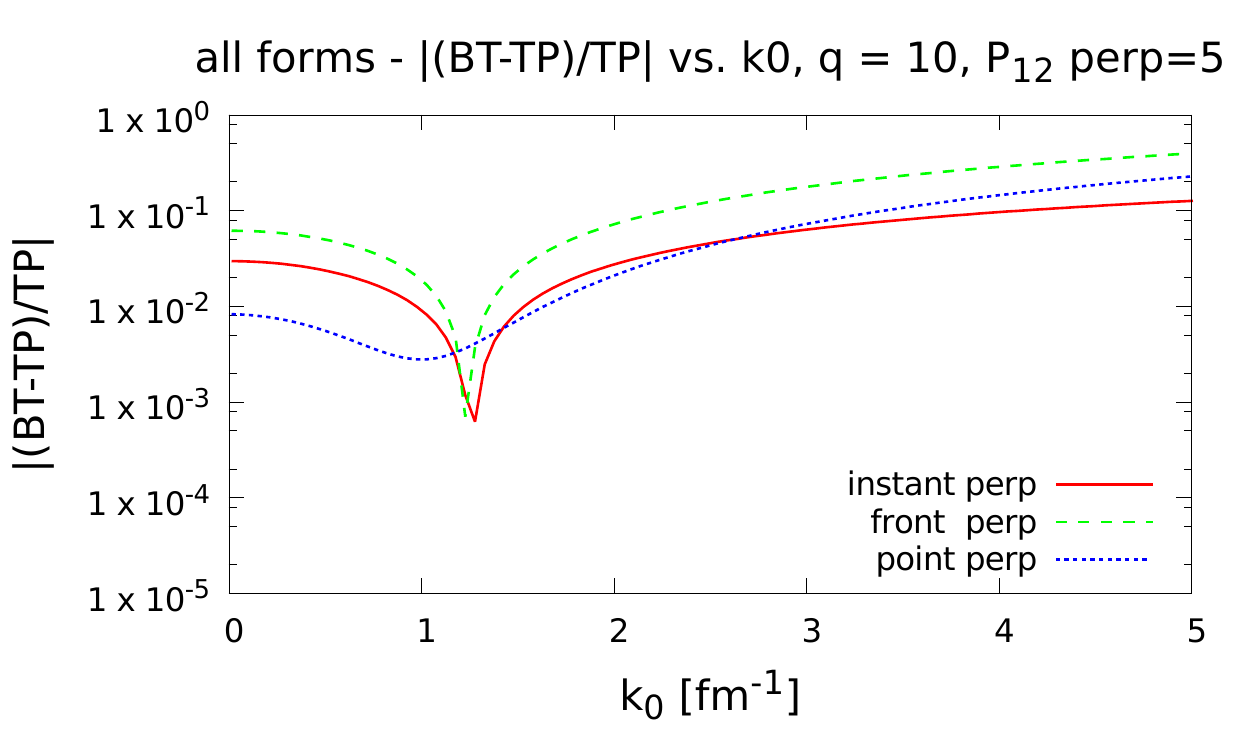}
    \caption{(Color online) Model differences for $BT$ calculations as a function of
      Gaussian wave function scale, $\q=10~{\rm fm}^{-1}\perp\p_{12}$.}
    \label{fig:quark-k0-all-prp}
  \end{minipage}
\end{figure*}

\section{Summary}

Bakamjian-Thomas formulations, which explicitly satisfy the
requirements of Poincar\'e invariance, do not satisfy cluster
separability above the three-particle level, i.e.\ in systems that
involve three-body subsystems whose total momentum must vary.  The
cluster properties can be restored via a hierarchy of unitary
transformations.  These transformations depend upon the full solution
of a three-body problem, and are difficult to implement in practice.

Rather than attempt to calculate directly the size of these unitary
transformations (e.g.\ the difference of matrix elements from those of
the unit operator), we have developed a simple model in which the
exact result consistent with cluster separability is known, and then
compare to it the results of Bakamjian-Thomas calculations.  The
quantities that we investigate are sensitive to the part of these
unitary transformations that restore cluster properties.

We conclude from these model studies that Bakamjian-Thomas models,
which explicitly satisfy the requirements of Poincar\'e invariance,
can be utilized in typical nuclear physics problems with minimal
quantitative error due to the lack of cluster separability.

We also examined models utilizing mass/momentum scales appropriate for
quark models.  Confinement precludes separating arbitrary subsystems
by large distance scales, so the general requirement of cluster
separability is irrelevant for models of individual hadrons.  The
issue may be relevant, however, for systems of hadrons described by
subnucleonic degrees of freedom.  In such cases, the deviations from
the model benchmark are larger (about 10\%) than those for typical
nuclear physics calculations with nucleons, though they are still
small compared to model uncertainties.

This work was supported in part by the U. S. Department of Energy, 
Office of Nuclear Physics, under contract No. DE-FG02-86ER40286.


\end{document}